\documentclass[aps,pre,floatfix,twocolumn, 
nofootinbib,groupedaddress]{revtex4}
\usepackage{graphicx}
\begin{document}
\title{Squeezed states and quantum chaos}
\author{Kirill N. Alekseev}
\altaffiliation[Current address: ]{Division of Theoretical Physics,
Department of Physical Sciences, Box 3000, University of Oulu FIN-90014,
Finland}
\author{Dima S. Primak}
\address{Theory of Nonlinear Processes Laboratory,
Kirensky Institute of Physics,\\
Russian Academy of Sciences,
Krasnoyarsk 660036, Russia}

\begin{abstract}
We examine the dynamics of a wave packet that initially corresponds to a
coherent state in the model of quantum kicked rotator. This main model of
quantum chaos, which allows for a transition from regular to to chaotic
behavior in the classical limit, may be realized in experiments with cold atoms.
We study the generation of squeezed states in the quasiclassical limit and in a
time interval when quantum-classical correspondence is yet well-defined. We find
that the degree of squeezing depends on the degree of local instability in the
system and increases with the Chirikov parameter of stochasticity. We discuss
the dependence of the degree of squeezing on the initial width of the packet,
the problems of stability and observability of squeezed states at the transition
to quantum chaos, as well as the dynamics of wave packet destruction.
\end{abstract}
\maketitle
\section{Introduction}
\label{sec:Intro}

At present the problem of squeezed quantum states generation draws a lot of
attention, both from the standpoints of
pure knowledge and possible applications \cite{reviews,reynaud,fabre}.
Most often the topic is squeezed states of the electromagnetic field. If in
the simplest case we consider a single-mode quantum field, which is described
by the creation $a^{\dag}$ and annihilation $a$ operators,
the variances of quadrature field operators $a_1=a+a^\dag$ and
$a_2=-i (a-a^\dag)$
satisfy the uncertainty relation $\Delta a_1 \Delta a_2\ge 1$,
where the equality holds for a coherent state or vacuum. Then, in these simple
terms, a squeezed state is a state for which the variance of one of the
quadrature components is less than unity. Quantum fluctuations, determined
by the uncertainty relation, can be represented diagrammatically in the
$a_1 - a_2$ plane by a circle for a coherent state or by an ellipse for a
squeezed state. In a more systematic description of squeezing, the
quantum-noise ellipse is determined in terms of the projection onto the
$a_1 - a_2$ plane of the horizontal section of the
Wigner distribution function, which gives the quasiprobability distribution
for measuring  the quadrature field components \cite{fabre}.
\par
A typical situation in experiments on generation of squeezed states
is one in which a large number of photons
participate in a nonlinear interaction and the amplitude of quantum
fluctuations is small compared to the averaged values
of the observables \cite{reynaud,fabre}. In this case the common approach in explanation
of squeezing is to use the semiclassical setting, where the Wigner quantum
function is actually replaced by a classical distribution function and
instead of examining the dynamics of the quantum-noise ellipse one considers
the evolution of the classical phase volume \cite{fabre,heidmann85}.
\par
For quite a long time it has been known that squeezing is enhanced if
a system is close to a bifurcation point between two different dynamical
regimes \cite{fabre,heidmann85,lugiato,heidmann85a}.
The increase of squeezing in such conditions was considered, e.g., for the
parametric interaction of light waves \cite{lugiato} and for the interaction
of Rydberg atoms with an electromagnetic mode in a high-$Q$
cavity in a dynamical regime close to the separatrix
\cite{heidmann85,heidmann85a}.
The following simple argument is used to explain the increase of squeezing
near a bifurcation point. Quantum fluctuations build up for the physical
variable that is unstable near the threshold. As a result there is nothing
to stop the strong squeezing for the conjugate variable because of
conservation of a phase volume in a nondissipative system \cite{fabre}.
\par
It must be noted at this point that a number of researchers (see Refs.
\cite{fabre,heidmann85,lugiato,heidmann85a}) studied the increase of squeezing
near the instability threshold in different optical systems with \textit{
regular dynamics}. However, it is well known that strong (exponential)
deformation of the phase volume is one of the main manifestations of dynamical
chaos in classical systems \cite{sagdeev-book}. The
physical reason for such strong deformations of the phase volume is the local
instability of chaotic motion, which usually
manifests itself within a wide range of values of the control parameter
of dynamical system.
According to the correspondence principle, in the quasiclassical limit
a quantum system must
mimic the properties of a classical system. Thus, it is quite natural
to expect the increase of squeezing at the transition to
quantum chaos. On the other hand, in a quantum mechanical description we speak
only of the dynamics of wave packets, whose center moves almost along a
classical trajectory in the course of a certain time interval.
Hence, the strong deformations of the phase volume,
which accompany the transition to chaos, must
manifest themselves in squeezing along a certain directions in phase space.
\par
As far as we know, the generation of squeezed states in systems with chaotic
dynamics was first examined in
Refs. \cite{alekseev-preprint,alekseev95,chirikov-conf}. By employing the
$1/N$-expansion method \cite{heidmann85a,yaffe} (here $N$ is the number of
quantum states participating in the dynamics of the system) it was found in
Refs. \cite{alekseev-preprint,alekseev95} that the squeezing of light increases
significantly at the transition to chaos
during the time interval for which quantum-classical correspondence is
well-defined \cite{berman-berry}. This result was first obtained in
Refs. \cite{alekseev-preprint,alekseev95} for the generalized $N$-atoms
Janes-Cummings model, which allows
a transition from regular to chaotic dynamics in the limit $N\rightarrow\infty$
\cite{alekseev88-94}. Later the increase of squeezing at transition to chaos
was found for arbitrary single mode quantum optical systems
\cite{alekseev97-98}.
The squeezing of wave packets in conditions of quantum chaos was also briefly
discussed in \cite{chirikov-conf}.
However, the main results of works
\cite{alekseev-preprint,alekseev95,alekseev97-98} were obtained by using a
form of perturbation theory (the $1/N$-expansion).
Therefore it is important to investigate the generation of
squeezed states at transition to quantum chaos in a simple quantum system
whose time evolution can be found explicitly.
\par
In the present paper we study the generation of squeezed states in the
time evolution of an initially Gaussian wave packet in the model of quantum
kicked rotator.
This model was first introduced by Casati et al. \cite{casati79} and at present
is the main model in studies of quantum chaos (for review see
\cite{chirikov81,15,casati95}). The quantum rotator model is attractive mainly
for two reasons: (i) the classical limit for
this model is a well-known standard map \cite{chirikov79}, and (ii)
it is fairly easy to study numerically the dynamics of the model
for a large number of quantum levels.
\par
We examine the dynamics of narrow Gaussian packets in a rotator with
$2^{17} (\simeq 10^5)$ levels. We define squeezing for the
generalized quadrature operator $X_\theta=a \exp(-i\theta)+a^\dag\exp(\theta)$,
where $\theta$ is a real parameter. It is just $X_\theta$ that is observed
in the homodyne detecting scheme, where $\theta$ is
determined by the phase of the reference beam \cite{reviews}. We will see that
as long as the wave packet is localized, the degree of
squeezing correlates well with the degree of local instability in the system.
Here the greater the instability, the stronger
the squeezing achieved in a shorter time interval. Squeezing is much stronger
for quantum chaos than it is for regular motion. We will also see that the
narrower the initial wave packet, the higher the degree of squeezing that
can be achieved.
We attribute this to the fact that a narrow wave packet is closer in its
time evolution to the classical trajectory than a broad
one, with the result that it is more sensitive to local instabilities in
the motion, which leads to stronger squeezing.
\par
We will also consider the problem of stability and observability of squeezed
states in the transition to chaos. More precisely,
we will study the time dependence of the optimum values of the phase
$\theta$ of the generalized quadrature operator $X_\theta$ for
which the squeezing is maximal (so-called \textit{principal squeezing}
\cite{luks,tanas}).
We will show that in conditions of strong chaos and for
long enough time, the optimum values of the phases change
dramatically even under a small perturbation of the
parameters of the initial Gaussian wave packet. Such a squeezing regime
is unstable and difficult to observe. On the other
hand, our results suggest that for weak chaos squeezing is fairly stable.
\par
We will also briefly discuss the dynamics of disintegration of wave packets
at chaos. Here we will show that a typical
scenario of disintegration of an initially localized
wave packet in chaos consists of two stages: the initial spread of the wave
packet, and the catastrophic disintegration of the
packet into many small subpackets. Here our results agree on the whole with
the results of Casati and Chirikov \cite{casati95}.
\par
Note that earlier the dynamics of narrow Gaussian wave packets
in the quasiclassical limit was studied numerically for the
models of a quantum kicked rotator \cite{fox1,lan},
quantum kicked top \cite{fox2}, and quantum
Arnold cat \cite{elston-preprint} in connection with the problem of
quantum-classical correspondence at quantum chaos. However, in
these papers the generation of squeezed states was not considered.
\par
The model of a quantum kicked rotator is very popular in theoretical studies of
quantum chaos. On the other hand, recently possibilities of implementing
variants of this model in optical systems have been discussed
\cite{stand-opt-exp}. Moreover, the model of
quantum rotator  has been realized in experiments on
the interaction of laser beam and cooled atoms \cite{atom-opt-exp}.
Hence our results on the increase of squeezing at the transition to quantum
chaos
in a kicked quantum rotator are also directly related to experimentally
realizable systems.
\par
The plan of the paper is as follows. In Sec. \ref{sec:Model}
we discuss the quantum standard map and find how to calculate principal
squeezing. The method used in numerical simulations is developed in
Sec. \ref{sec:Numerical_method}, and the main results on the
dynamics of squeezing are given in Sec. \ref{sec:Results}. Finally,
in Sec. \ref{sec:Disc} we draw the main conclusions and consider the
possibility of verifying our results in experiments.

\section{The model of quantum kicked rotator and squeezed states}
\label{sec:Model}

Consider the model of a quantum rotator with periodic delta-kicks. Here we
follow the notation of Ref. \cite{lan}.
The Hamiltonian is
\begin{eqnarray}
\label{1}
H &= &\frac{p^2}{2 m L^2}-\delta_p(t/T) m L^2 \omega_0^2 \cos{(x)},\\
& &\delta_p(t/T) = \sum_{j=-\infty}^{+\infty} \delta(j-t/T),\nonumber
\end{eqnarray}
where $x$ is the cyclic variable with a period $2\pi$, $L$ is the
characteristic size of the rotator, $m$ is the rotator mass, and $\omega_0$
is the frequency of linear oscillations. The function $\delta_p(t/T)$
describes a periodic sequence of kicks with a period $T$, where
$\delta(x)$ is the Dirac $\delta$-function. Let us introduce new variables
\begin{equation}
\label{2}
\alpha = m L^2 \omega_0^2 T, \quad \beta = \frac{T}{m L^2}
\end{equation}
and measure time in units of $T$, i.e., $ t \to t/T $. Then the Schr\"{o}dinger
equation takes the form
\begin{equation}
\label{3}
i\hbar\frac{\partial\Psi}{\partial t} =
 -\frac{\hbar^2\beta}{2}\frac{\partial^2\Psi}{\partial x^2}
 -\delta_p(t)\alpha\cos{(x)} \Psi.
 \end{equation}
Due to the periodicity of $\Psi(x)$ in $x$ the solution of Eq.~(\ref{3})
can be written as follows
\begin{eqnarray}
\label{4}
\Psi(x) &=& \frac{1}{\sqrt{2\pi}}\sum_{k=-\infty}^{+\infty} e^{i k x} A_k(t),
\nonumber \\
A_k(t)&=& \frac{1}{\sqrt{2\pi}}\int_0^{2\pi}\Psi(x) e^{-i k x} dx.
\end{eqnarray}
Using the standard procedure \cite{chirikov81,lan}, we obtain the quantum map
in the form
\begin{eqnarray}
\label{5}
\Psi_{n+1}&=&U_x U_p \Psi_{n}, \\ 
& &U_p=\exp{\left(-\frac{i\beta}{2\hbar}\hat{p}^2\right)},\quad
 U_x=\exp{\left(\frac{i\alpha}{\hbar}\cos{(\hat{x})}\right)},\nonumber
\end{eqnarray}
where $\Psi_n$ is the value of the wave function at the time moment
immediately after the $n$-th kick. The time evolution of the wave
function in the map (\ref{5}) is determined solely by two parameters,
$\alpha/\hbar$ and $\beta\hbar$.
Since $U_p$ is diagonalized in the $p$-representation,
$U_x$ is diagonalized in the $x$-representation, and the transition between
$x-$ and $p$-representations is given
by the Fourier transformation (\ref{4}), the map (\ref{5}) actually reduces to
\begin{equation}
\label{6}
\Psi_{n+1}(x)=U_x F^{-1} U_p F \Psi_n(x),
\end{equation}
where $F$ and $F^{-1}$ are the direct and inverse Fourier transforms.
\par
Sometimes it proves useful to use the quantum map written in terms of
the probability amplitudes $A_k$ of transitions
between the unperturbed levels of the rotator \cite{casati79}. Combining
Eqs. (\ref{4}) and (\ref{5}), we obtain
\begin{eqnarray}
\label{6'}
A_k^{(n+1)}&=&\sum_{m=-\infty}^{+\infty} F_{km} A_m^{(n)},\\
& &F_{km}=(-i)^{k-m} \exp{\left(-\frac{i\hbar\beta m^2}{2} \right)}
J_{k-m}\left(\frac{\alpha}{\hbar}\right), \nonumber
\end{eqnarray}
where $J_l(z)$ is the Bessel function of order $l$ and argument $z$,
and the superscript $(n)$ on the variable $A$ stands for the
number of the kick. Bearing in mind that the Bessel functions with
$\mid l \mid\geq z$ rapidly decrease with increasing $l$, we see from Eq.
(\ref{6'}) that, with exponential accuracy, in the course of a single kick
$2\alpha{\hbar}^{-1}$ unperturbed rotator levels are captured. Below
we consider the case where $\alpha/\hbar$ is large, which is typical of
quantum chaos problems.
\par
\begin{figure}
\includegraphics[width=1\linewidth]{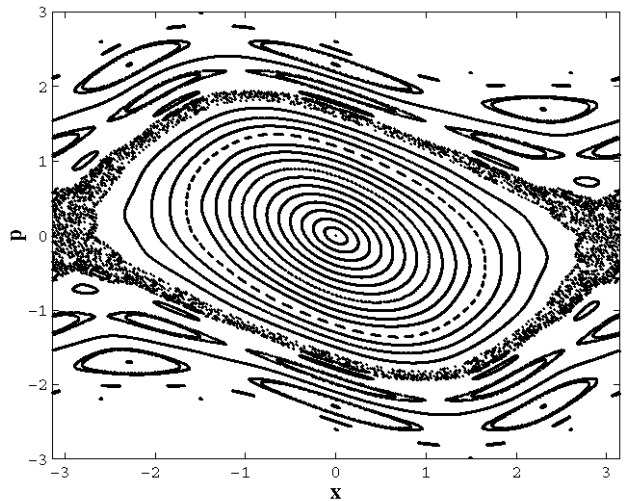}
\caption{Phase portrait of the classical standard map for $K=0.8$.}
\label{fig1ps}
\end{figure}

In the classical limit the Hamiltonian (\ref{1}) reduces to the standard map
\begin{equation}
\label{7}
P_{n+1}=P_n-K\sin x_{n+1}, \quad
x_{n+1}=x_n+P_n \pmod{2\pi},
\end{equation}
where $P_n=\beta p_n$, the subscript $n$ denoting the values of $x$ and $P$
immediately after the $n$-th kick, and $K\equiv\alpha\beta$ is the
Chirikov parameter\footnote{
The unusual form of the standard map (\ref{7}) is due to the fact that we find
it convenient to take the values of $x$ and $p$ immediately after
the $n$-th kick rather than before the $n$-th kick as is done in majority
of papers. It must be noted however that nonlinear dynamics is, of course,
same for both these forms of the standard map.
}.
Strong and global chaos sets in for $K>1$. For $K< 1$ the larger part of the
phase plane is filled
with regular trajectories, although small regions with local chaos exist no
matter how small $K$ may be \cite{chirikov79}. The phase
portrait for the map (\ref{7}) at $K=0.8$ is depicted in Fig.~\ref{fig1ps}.
The chaotic layer lies near the separatrix of the main resonance,
which passes through the hyperbolic points $(\pm\pi,0)$. In our calculations
we usually take a wave packet whose center
is located near a hyperbolic point but yet in a region of regular motion at
$K<1$.
\par
For the initial state of the quantum map (\ref{5}) we take the Gaussian wave
packet
\begin{equation}
\label{8}
\Psi(x)=(2\pi\sigma^2)^{-1/4}
 \exp{\left(-\frac{(x-x_0)^2}{4\sigma^2}+i k_0 (x-x_0)\right)} ,
\end{equation}
where $\langle x\rangle=x_0$, $\langle\delta x^2\rangle\equiv
\langle x^2\rangle-\langle x\rangle^2=\sigma^2$,
$p_0\equiv\langle p\rangle=\hbar k_0$,
$\langle\delta p^2\rangle=\frac{\hbar^2}{4\sigma^2}$ and $k_0$ is an integer.
The packet is assumed narrow: $\sigma\ll x_0$ and
$\langle \delta p^2 \rangle \ll \hbar k_0$.
Note that in view of its periodicity in $x$ the wave packet (\ref{8})
is generally not a state that minimizes the uncertainty relation. But in the
case of a narrow packet it is essentially indistinguishable from a
minimum-uncertainty state \cite{fox1,lan,fox2}.
\par
A typical initial quantum state in studies of light squeezing is a coherent
state  \cite{reviews,reynaud,fabre}. Such a state is an eigenfunction of the
annihilation operator $a$, which in the present notation can be written as
\begin{equation}
\label{9}
a=\frac{1}{\sqrt{2\hbar}}(\sqrt{\gamma}\hat{x}+i\frac{\hat{p}}{\sqrt{\gamma}}),
\quad \gamma={\left(\frac{\alpha}{\beta}\right)}^{1/2}.
\end{equation}

The fact that the annihilation operator has such an appearance can easily
be understood if we consider the following
limiting case of the harmonic oscillator that follows from (\ref{3}):
\begin{equation}
\label{10}
i\hbar\frac{\partial\Psi}{\partial t} =
    -\frac{\hbar^2\beta}{2}\frac{\partial^2\Psi}{\partial x^2}
    +\frac{\alpha x^2}{2} \Psi.
\end{equation}
Now we can show that the wave function (\ref{8}) is a coherent state,
i.e., an eigenfunction of operator (10), if we put
\begin{equation}
\label{11}
\sigma^2=\frac{\hbar}{2\gamma}.
\end{equation}
\par
Let us now turn to the problem of squeezing.
In light squeezing experiments \cite{reviews}, the observable quantity is the
variance of the generalized quadrature operator
\begin{equation}
\label{12}
X_\theta=a e^{-i\theta}+a^\dag e^{i\theta},
\end{equation}
where $\theta$ is the phase of the reference beam in the homodyne
detecting scheme. In the particular cases where $\theta=0$ and
$\theta=\pi/2$,
Eq.~(\ref{12}) yields the following expressions for the generalized position
and momentum operators
\begin{equation}
\label{12'}
X_1=a+a^\dag,\quad X_2=-i(a-a^\dag),\quad [X_1,X_2]=2i
\end{equation}
with the uncertainty relation
$\langle\delta X_1^2\rangle \langle\delta X_2^2\rangle\geq 1$,
where averaging is done over an arbitrary quantum state and equality is
achieved for a coherent state. The standard definition of quadrature
squeezing is the condition \cite{reviews,fabre}
\begin{equation}
\label{14}
\min{\left(\langle\delta X_1^2\rangle, \langle\delta X_2^2\rangle\right)}<1,
\end{equation}
i.e., the variance of one of the quadrature components is smaller than for the
coherent state.
\par
In a more general case we consider the variance
$\langle\delta X_\theta^2\rangle$
of the operator (\ref{12}), and the state is assumed squeezed if the
value of $\langle\delta X_\theta^2\rangle$
in this state for some value of $\theta$ is smaller than in the coherent state
\cite{luks,tanas}. Experiments actually determine
the minimum $S$ of this variance as a function of the angle $\theta$ as
\begin{equation}
\label{15}
S=\min_{\theta \in [0,2\pi]}{\langle\delta X_\theta^2\rangle}.
\end{equation}
Using the definition (\ref{12}) of $X_\theta$, it is possible to show
\cite{luks,tanas} that
\begin{equation}
\label{16}
 S=1+2\langle\delta a^\dag \delta a\rangle-
 2\sqrt{\langle\delta a^2\rangle\langle\delta a^{\dag 2}\rangle}
\end{equation}
and the minimum of $\langle\delta X_\theta^2\rangle$ is reached at an optimum
phase value $\theta=\theta^*$ defined as follows \cite{tanas}:
\begin{equation}
\label{17}
e^{2i\theta^*}=-\sqrt{\langle\delta a^2\rangle/\langle\delta a^{\dag 2}\rangle}
\end{equation}
For our discussion it is convenient to express S in terms of the cumulants
of the operators $x$ and $p$. Using the definition  (\ref{9})
of operator $a$ and Eq.~(\ref{16}), we obtain
\begin{equation}
\label{18}
 S=\frac{1}{\hbar}\left(\frac{\langle\delta p^2\rangle}{\gamma}+
 \langle\delta x^2\rangle\gamma
-\sqrt{(\langle\delta x^2\rangle\gamma-\langle\delta p^2\rangle/\gamma)^2+
4c^2}\right),
\end{equation}
where $c=\frac{1}{2}(\langle (x p+p x)\rangle-
2\langle x\rangle\langle p\rangle)$
Clearly for any  Gaussian wave packet $S=\frac{\hbar}{2\sigma^2\gamma}$,
while for a coherent state we have $S=1$ because of equality (\ref{11}).
Hence a state is squeezed if
\begin{equation}
\label{19}
S<1
\end{equation}
The condition determines the principal
squeezing attainable in homodyne detection \cite{luks}.
The maximum of the variance $\langle\delta X_\theta^2\rangle$ in $\theta$
can be defined in the same way the
minimum was defined in Eq.~(\ref{15}). We denote
it by $\overline{S}$. Then we can show that the dependence of $\overline{S}$
on the cumulants differs from Eq.~(\ref{18}) only in the sign in front of the
square root, so that we have
\begin{equation}
\label{20}
S\overline{S}\geq 1.
\end{equation}
Thus, squeezing in $S$  (Eq.~\ref{19}) is accompanied by dilation
in $\overline{S}$.
Note that in contrast to the quadrature squeezing (\ref{14}), the definition
of principal squeezing (\ref{18}) contains quadrature
correlators of the $\langle x p\rangle$ type. This is very important for
systems with discrete time, to which the model of a quantum
kicked rotator belongs. The thing is that the quadrature squeezing
(\ref{14}) is essentially unobservable in such systems, although the principal
squeezing (\ref{18}) and (\ref{19}) may occur
\footnote{
Note that Lan \cite{lan} have studied the time dependence of
$\langle\delta q^2\rangle$ and $\langle\delta p^2\rangle$
for a quantum kicked
rotator on a time scale on which one-to-one quantum-classical
correspondence holds (see Table I in Ref. \cite{lan}). Both these variances
increase uniformly so that quadrature squeezing (\ref{14}) is impossible
}.
We will discuss the time dependence of $S$ in Sec.~\ref{sec:Results}.

\section{The numerical method}
\label{sec:Numerical_method}

Several features of the numerical method must be mentioned. The interval in
$x$ from $0$ to $2\pi$ is partitioned into $N$
segments $\triangle x = 2\pi/N$, and the wave function $\Psi(x)$ is represented
by a discrete sequence of values (column vector $\mid\Psi\rangle$) of
length $N$, so that $\Psi_l=\Psi(l\triangle x)$, $l\in [0,1,\ldots,N-1] $.
Accordingly, $k$ varies from $0$ to $N-1$ in the sum (\ref{4}).
In our numerical
method $N$ is always an integral power of two
and the operator $F$ in Eq.~(\ref{6}) is represented in the form
of the fast Fourier transform inducing the transformations
\begin{equation}
\label{21}
 F:\Psi_l\to A_k,\quad F^{-1}:A_k\to\Psi_l
\end{equation}
To determine the principal squeezing, we must calculate
$\langle\delta q^2\rangle$, $\langle\delta p^2\rangle$, and
$\langle x p\rangle$ (see Eq.~(\ref{18})). For instance, the
calculation of $\langle x p\rangle$ proceeds along the following lines
$$
\langle  x p\rangle = \langle\Psi|
\vec{x} F^{-1} \vec{p} F |\Psi\rangle,
$$
where $\langle\Psi|$ is obtained by transposing the vector $|\Psi\rangle$
and then finding the complex conjugate of the result, while
$\vec{x}$ and $\vec{p}$ are vectors that initially have the form
$\vec{x}=[0,\triangle x,2\triangle x,\ldots,2\pi-\triangle x]$,
$\vec{p}=[0,1,2,\ldots,N-1]$.
\par
The fact that $x$ is defined in $\pmod{2\pi}$ requires following the wave
packet and ensuring that it is defined correctly during the passage through
the end-points of the interval $[0,2\pi]$. We set up the process in the
following manner. When
the center of the wave packet in the $x$-representation approaches an edge of
the half-interval $[0,2\pi[$, the wave function $\Psi(x)$
is examined on a new interval, $[-\pi,\pi[$, with a new vector
$\vec{x} = [0,\triangle x,\ldots,\pi,
    -\pi+\triangle x,-\pi+2\triangle x,\ldots,-2\triangle x,-\triangle x]$,
since $(-k\triangle x)\bmod{2\pi} = (2\pi-k\triangle x)\bmod{2\pi}$, where $k$
is an integer. The transition from $[-\pi,\pi[$ to $[0,2\pi[$ is treated
similarly.
\par
Calculations in the $p$-representation have their own special features.
Although for the Hamiltonian (\ref{1}) the
momentum is defined in the interval from $-\infty$ to $+\infty$,
in numerical calculations we deal only with a finite range of
values of momentum $p$, which is specified by the number terms $N$ in the
Fourier transform (\ref{4}).
To avoid the possible problem of reflection of the wave packet from an edge
of the given interval in the $p$-representation\footnote{
For a discussion of the problem of reflection and splitting of a wave packet
due to the finite range of momentum in the close model
of the quantum Arnold cat see Ref.~\cite{elston-preprint}.
}, we select this
interval in each iteration of map (\ref{6}) in such a way that the
maximum of the absolute value of the wave function of the
packet is always at the center of the given interval
(actually, we renumber the vector $\vec{p}$).
\par
The process of calculating the next iteration of the quantum map
(\ref{6}) is terminated as soon as the wave packet ceases to be
sufficiently localized either in the $x$-representation or in the
$p$-representation, i.e., when the number of terms in the Fourier
transformation actually involved in the calculation process is smaller
than needed.
Let us now to present the conditions for wave packet
delocalization, which we use in our numerical simulations.
First, introduce the notations
$\xi = \max_{[0,2\pi]}|\Psi(x)|$,
$\chi = \max\{|A_1|,|A_2|,\ldots,|A_N|\}$.
Next, introduce  $A_{left}$ ($A_{right}$),
the values of $A_k$ belonging, respectively, to the left and right edges
of the finite interval in which
the wave function in momentum space is determined.
The wave packet is delocalized and further calculations are
terminated when one of the two inequalities,
$$
\max\{|A_{left}|/\chi,|A_{right}|/\chi\} > \varepsilon 
\quad\mbox{or}\quad
|\Psi(z)|/\xi > \varepsilon 
$$
become valid (here $z=0$, if $x\in [0,2\pi [$ or
$z=\pi$ if $x\in [-\pi,\pi [$).  In this paper we used the cut off value
$\varepsilon = 0.002$.

\section{The main results}
\label{sec:Results}

As an initial wave function in our calculations we took the coherent state
(a Gaussian wave packet) with $\hbar=10^{-6}$, $k_0=10000$, and $\sigma$
was varied between 0.04 and 0.07.
We fixed the initial width of the wave packet $\sigma$ and the value of
Chirikov parameter $K$. Then, the parameters $\alpha$ and $\beta$
in the evolution operator (\ref{5}) are
\begin{equation}
\label{22}
\alpha=K^{1/2}\frac{\hbar}{2\sigma^2},\quad
\beta=K^{1/2}\frac{2\sigma^2}{\hbar}.
\end{equation}
These formulas are obtained by combining the definition $K=\alpha\beta$
and Eqs.~(\ref{9}) and (\ref{11}).
In Sec.~\ref{sec:Model} we found that the number of unperturbed
levels of rotator captured in
one kick is $\approx 2\alpha/\hbar$. From (\ref{22}) it
follows that in our case this number is $K^{1/2}/\sigma^2$ and amounts to
several tens of thousands for the adopted widths of the
wave packet.
\par
In our calculations $K$ was varied between 0.2 and 2 with a step of 0.02.
We found the time dependence of the squeezing
$S$ (\ref{18}) and the optimum value of the phase $\theta^*$ at which
$\langle\delta X_\theta^2\rangle$ is at its minimum. To demonstrate the
correlation that
exists between the degree of squeezing and characteristics of chaos
\cite{alekseev95,alekseev97-98} we calculated
\begin{equation}
\label{23}
d = [\langle\delta x^2\rangle+\langle\delta p^2\rangle]^{1/2}.
\end{equation}
It can be shown \cite{alekseev95,alekseev97-98,sundaram}
that in the classical limit and while the wave packet is well-localized, i.e.,
$[\langle\delta x^2\rangle]^{1/2}\ll x_0$  and
$[\langle\delta p^2\rangle]^{1/2}\ll p_0$,
the $d$ of (\ref{23}) corresponds to the following distance in phase space
\begin{equation}
\label{24}
d_{cl}(t) = [(\Delta x)^2+(\Delta p)^2]^{1/2},
\end{equation}
where $\left(\Delta x(t),\Delta p(t)\right)$ is the solution of the linear
small-perturbation  equations  near the  classical  trajectory
$\left( x(t),p(t)\right)$.
The quantity $d_{cl}(t)$ characterizes the divergence of two initially close
trajectories and enters into the definition of
the largest classical Lyapunov exponent
\begin{equation}
\label{25}
\lambda=\lim_{t\rightarrow\infty}\frac{d_{cl}(t)}{t}.
\end{equation}
For a classical standard map in conditions of strong chaos, $K\gg 1$, there
exists the simple dependence $\lambda\approx\ln(K/2)$ \cite{chirikov79}.
Lyapunov exponent (\ref{25}) is an asymptotic characteristic of chaos.
For finite time intervals \cite{sagdeev-book}
\begin{equation}
\label{26}
d_{cl}(t)\simeq \exp(h(x,p) t),
\end{equation}
where the exponent $h$ is a function of a point in phase space and coincides,
in order of magnitude, with the Lyapunov
exponent $\lambda$, but in some time intervals the difference between the two
may be significant. The latter fact can be
explained by the strong inhomogeneity in the statistical properties of the
phase space of chaotic systems and,
correspondingly, by the different rates of divergence of trajectories in
different regions of phase space through which
the system passes in its time evolution. It must be noted at this point that
the dependence of $h$ on the parameter $K$ is
rather complicated. What is important, however, is only the property of the
strong (exponential) increase of $d_{cl}$ (\ref{26})
in the presence of chaos, a property often called
\textit{local instability} \cite{sagdeev-book}. When the motion is regular,
the time dependence of $d_{cl}$
is much weaker -- it follows a power function \cite{sagdeev-book}.
\par
On the other hand, it is $h$ that determines the rate of phase volume
deformation: the stronger the local instability, the
greater the deformation of phase volume in a given time interval.
Since in our case quantum-classical correspondence and the concept of chaos
are well-defined only in a very short time
interval, while the wave packet remains localized, it is meaningful to consider
the correlations existing between the
time dependence of the squeezing and that of the quantity $d$
(see (\ref{23})), which in the classical limit becomes $d_{cl}$ (see
(\ref{24})).
\par
\begin{figure}
\includegraphics[width=0.9\linewidth]{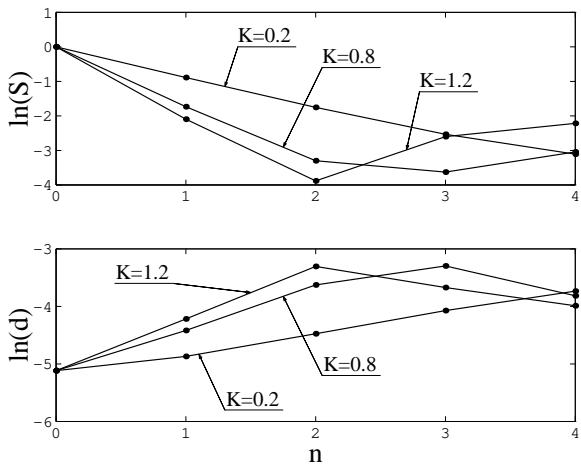}
\caption{Upper part of the figure:
Time dependence of the logarithm of squeezing $\ln S$.
Lower half of the figure:
Time dependence of $\ln d$ (see Eq.~(\ref{23})).
Everywhere $x_0=\pi$ and $\sigma=0.006$.}
\label{fig2ps}
\end{figure}

Figure~\ref{fig2ps} depicts the time dependence of the logarithm of squeezing
$\ln S$ and $\ln d$ for different values of $K$, when the
center of the wave packet is initially located
at the point $x_0=\pi$, $p_0=\hbar k_0=0.01$.
This initial condition is close to a
hyperbolic point through which the chaotic layer passes even when $K$ is small
(see Fig.~\ref{fig1ps})\footnote{We should note that for $K=0.8$
(Fig.~\ref{fig1ps}) this position of wave packet does not belong to chaotic
layer.
}.
Figure~\ref{fig2ps} shows that the
larger the squeezing (the smaller the value of $S$) the larger the local
instability (the larger the values of $\ln d$). This valid up to $n\approx 4$,
when the packet spread becomes so large that purely quantum effects become
important.
\par
\begin{figure}
\includegraphics[width=0.9\linewidth]{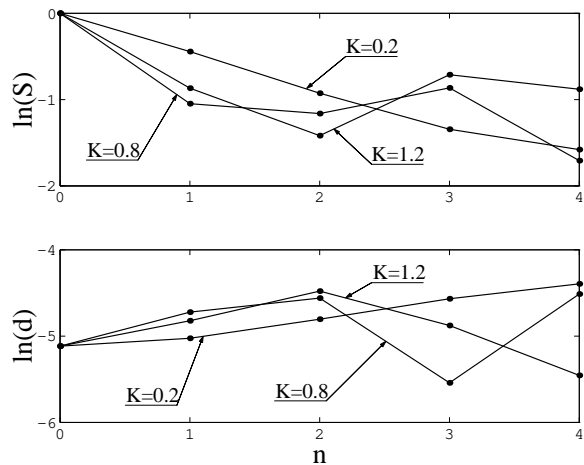}
\caption{Same as in Fig.~\ref{fig2ps} but for $x_0=\pi/2$.}
\label{fig3ps}
\end{figure}

For another initial condition, $x_0=\pi/2$, $p_0=0.01$,
which is closer to an elliptic point and hence lands in the chaotic
region only at large values of $K$, the dynamics of squeezing is depicted in
Fig.~\ref{fig3ps}.
We see that in this case squeezing is
in about two orders stronger than in the conditions of
Fig.~\ref{fig3ps} in the same time interval. On the
other hand, both Fig.~\ref{fig2ps} and Fig.~\ref{fig3ps} exhibit an increase
of squeezing as a function of the parameter $K$, which controls the
strength of chaos in the system.
\par
\begin{figure}
\includegraphics[width=0.8\linewidth]{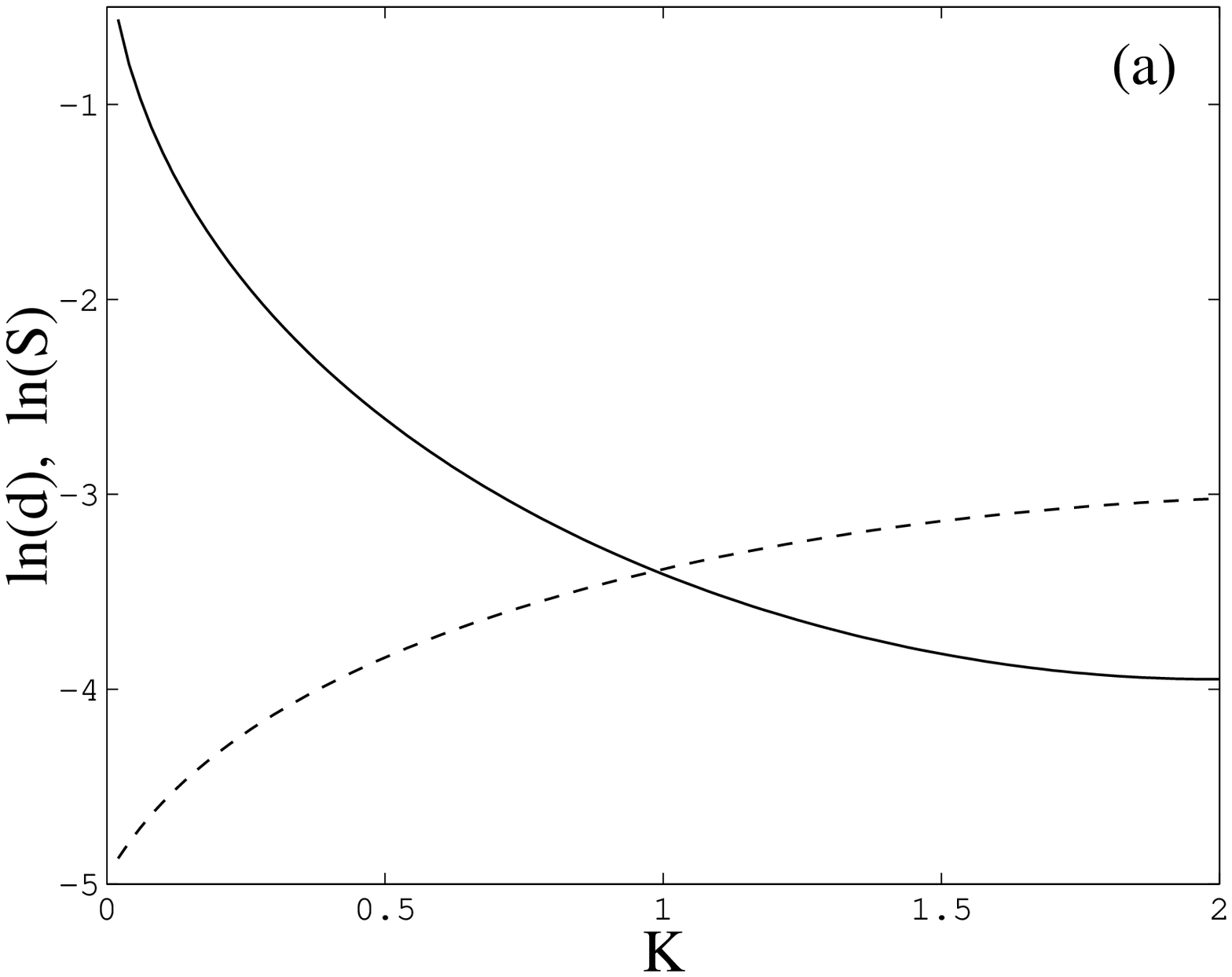}
\includegraphics[width=0.8\linewidth]{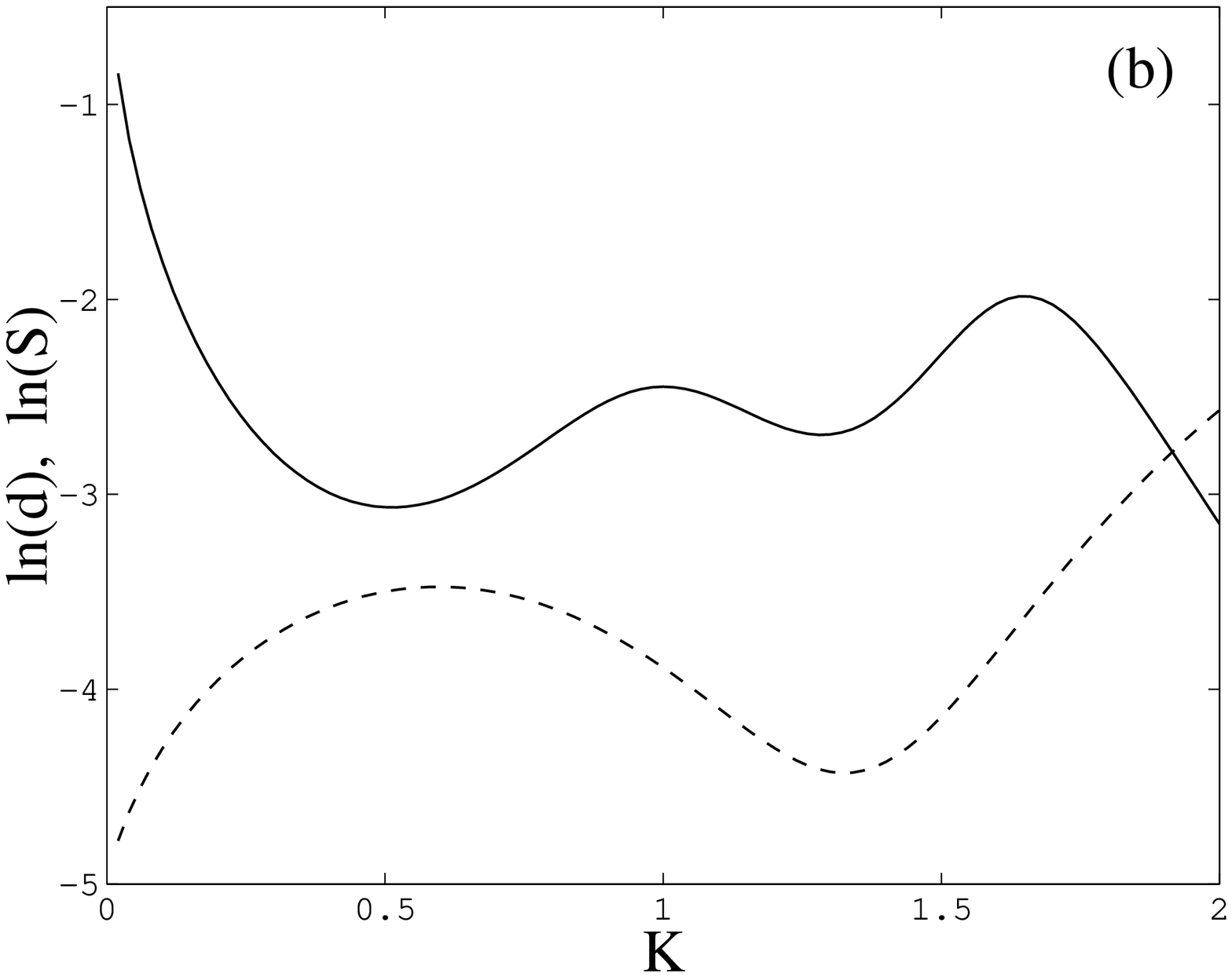}
\includegraphics[width=0.8\linewidth]{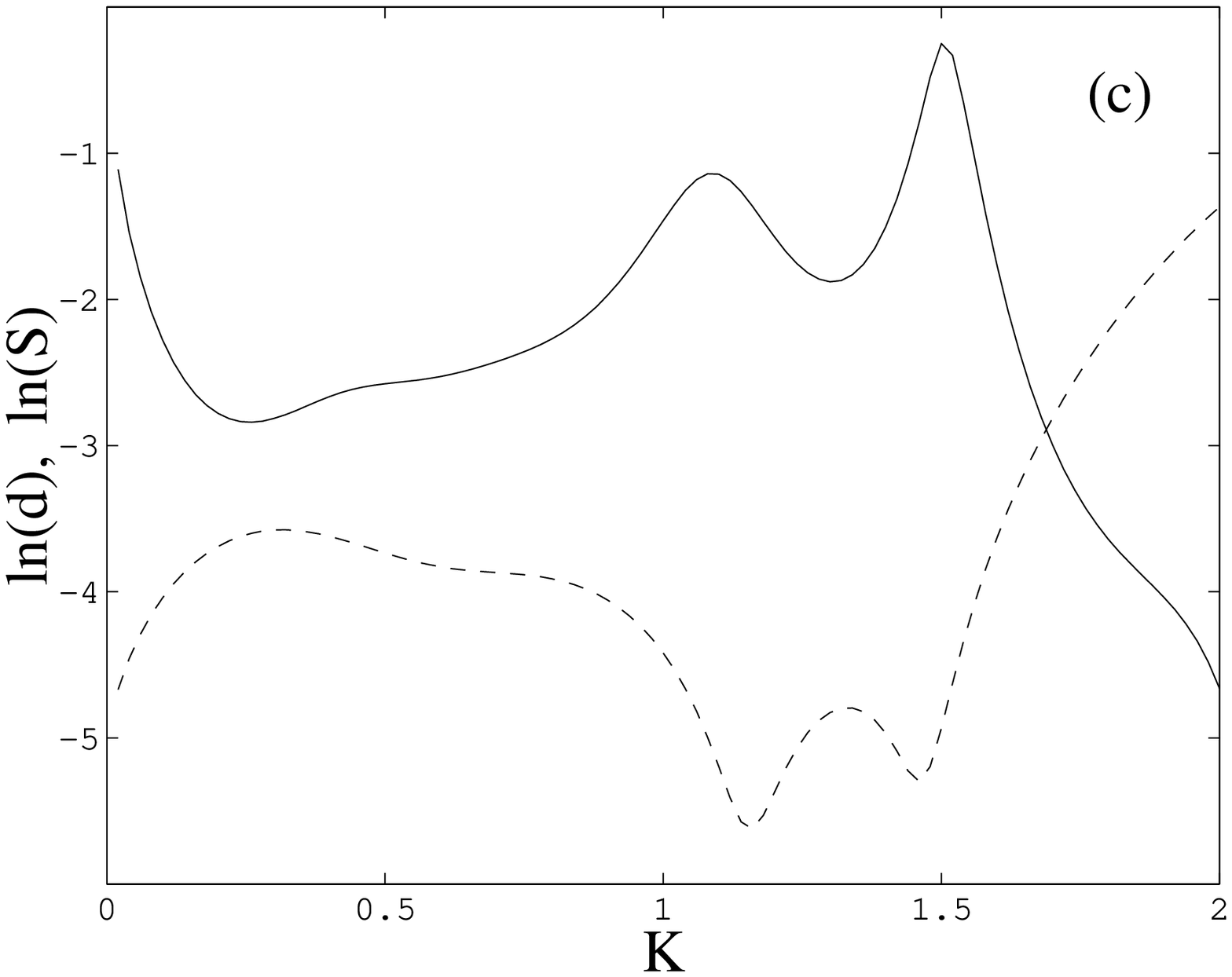}
\caption{The dependencies of $\ln S$ (solid curves) and $\ln d$ (dashed curves)
on the Chirikov parameter $K$ at a fixed number of kicks: (a) $n=3$, (b) $n=4$,
(c) $n=5$. Everywhere $x_0=\pi$ and $\sigma=0.007$.}
\label{fig4ps}
\end{figure}
\begin{figure}
\includegraphics[width=0.9\linewidth]{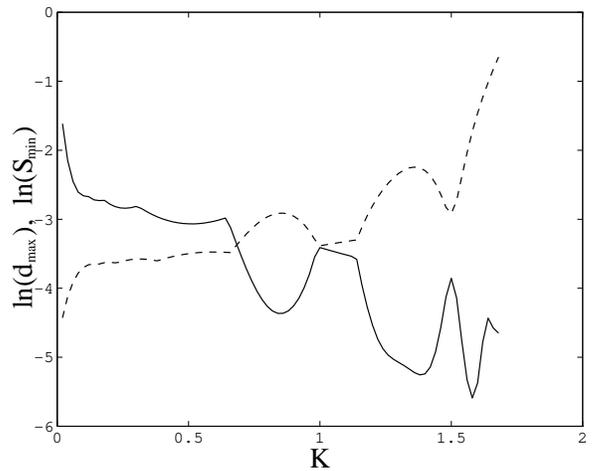}
\caption{Logarithm of the minimal squeezing $S_{min}$ (solid curve)
and logarithm of the maximal local instability $d_{max}$ (dashed curve)
as functions of the Chirikov parameter $K$ after seven kicks. Other parameters
and initial conditions are the same as in
Fig.~\ref{fig4ps}.}
\label{fig5ps}
\end{figure}

Now we turn to study of the correlation between squeezing and the degree of
local instability at different $K$ in greater detail.
The $K$-dependence of the degree of squeezing calculated after a fixed number
of kicks at $x_0=\pi$  and $p_0=0.01$
is depicted in Fig.~\ref{fig4ps}.
After the third kick the correlation between
$\ln S$ and $\ln d$ become very evident (Fig.~\ref{fig4ps}a).
However, small discrepancies in this dependence may appear as the number of
kicks grows. Such discrepancies become
evident, for instance, after the fourth kick for $1.1\lesssim K\lesssim 1.4$
(Fig.~\ref{fig4ps}b). After five kicks, $n=5$, the correlation between
$\ln S$ and $\ln d$ is restored (Fig.~\ref{fig4ps}c). Note that this
behavior pattern is quite typical. Hence, to establish the correlation
between local instability and squeezing more
clearly, a certain procedure of coarsening (averaging) these quantities in
the given time interval is needed. In our study
we determine the minimal squeezing, $S_{min}$,  and the maximal $d_{max}$
in a time interval during which the packet remains well localized for most
values of $K$ considered here. We found that there is a distinct
correlation between
$S_{min}$ and $d_{max}$: the larger the value of $d_{max}$ the smaller the
value of $S_{min}$, and vice versa. An example of such a dependence is
depicted in Fig.~\ref{fig5ps}, where $S_{min}$ and $d_{max}$
were calculated after six kicks.
Note that the diagrams do not go farther than $K>1.7$ because after six kicks
the wave packet becomes delocalized for $K>1.7$ and calculating averages and
local instability becomes meaningless.
\par
\begin{figure}
\includegraphics[width=0.9\linewidth]{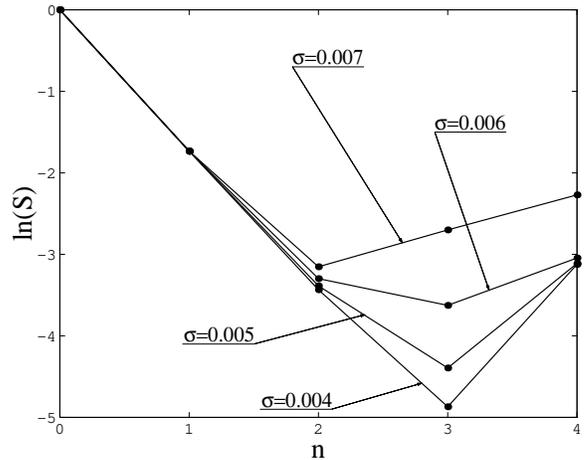}
\caption{Time dependence of the logarithm of squeezing $S$ for different
initial widths of the  wave packet $\sigma$ and at fixed values $K=0.8$,
$x_0=\pi$.}
\label{fig6ps}
\end{figure}

We also studied the dependence of the dynamics of squeezing on the initial
width of the wave packet $\sigma$. The results are
depicted in Fig.~\ref{fig6ps}.
Clearly, the narrower the packet the stronger
the squeezing achieved in a fixed time interval. This
dependence arises because a narrow wave packet travels farther along
its classical trajectory than a wide packet, so that
it undergoes stronger deformations related to nonlinear classical dynamics.
The exponential decrease is replaced by growth when the wave packet departs
from the classical limit and the dynamics is of an essentially
quantum nature.
\par
Now examine the problem of stability and observability of squeezing in
conditions of chaos. The figures mentioned earlier can
serve to illustrate the statement that the stronger the chaos the
stronger the principal squeezing. However, the definition
of principal squeezing (\ref{18}) is related to fixing the phase,
$\theta=\theta^*$. Here $\theta^*$ is time dependent even for exactly
integrable systems \cite{tanas}. For strong chaos in the classical limit,
the time dependence of $\theta^*$ may be very complicated.
Indeed, in addition to stretching and squeezing, the main feature of classical
chaos in the systems with bounded phase
space is the multiple formation of folds of the phase volume as a system
evolves \cite{sagdeev-book}. Hence the process of finding the
``minimum width'' of a phase drop, which actually amounts to finding
the dependence $\theta^*$ vs $t$ in the quasiclassical limit,
becomes unstable for large time intervals.
\par
\begin{figure}
\includegraphics[width=0.7\linewidth]{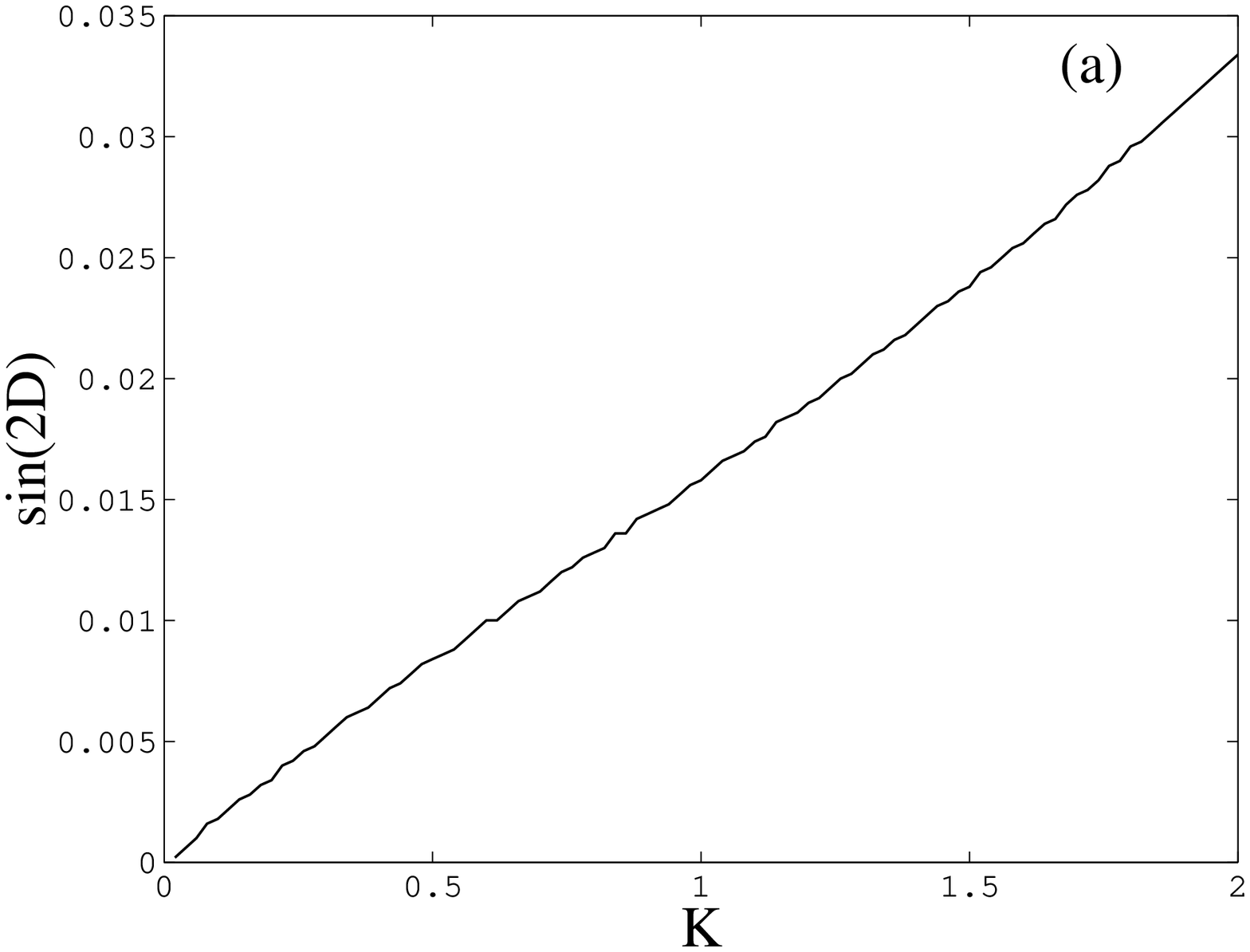}
\includegraphics[width=0.7\linewidth]{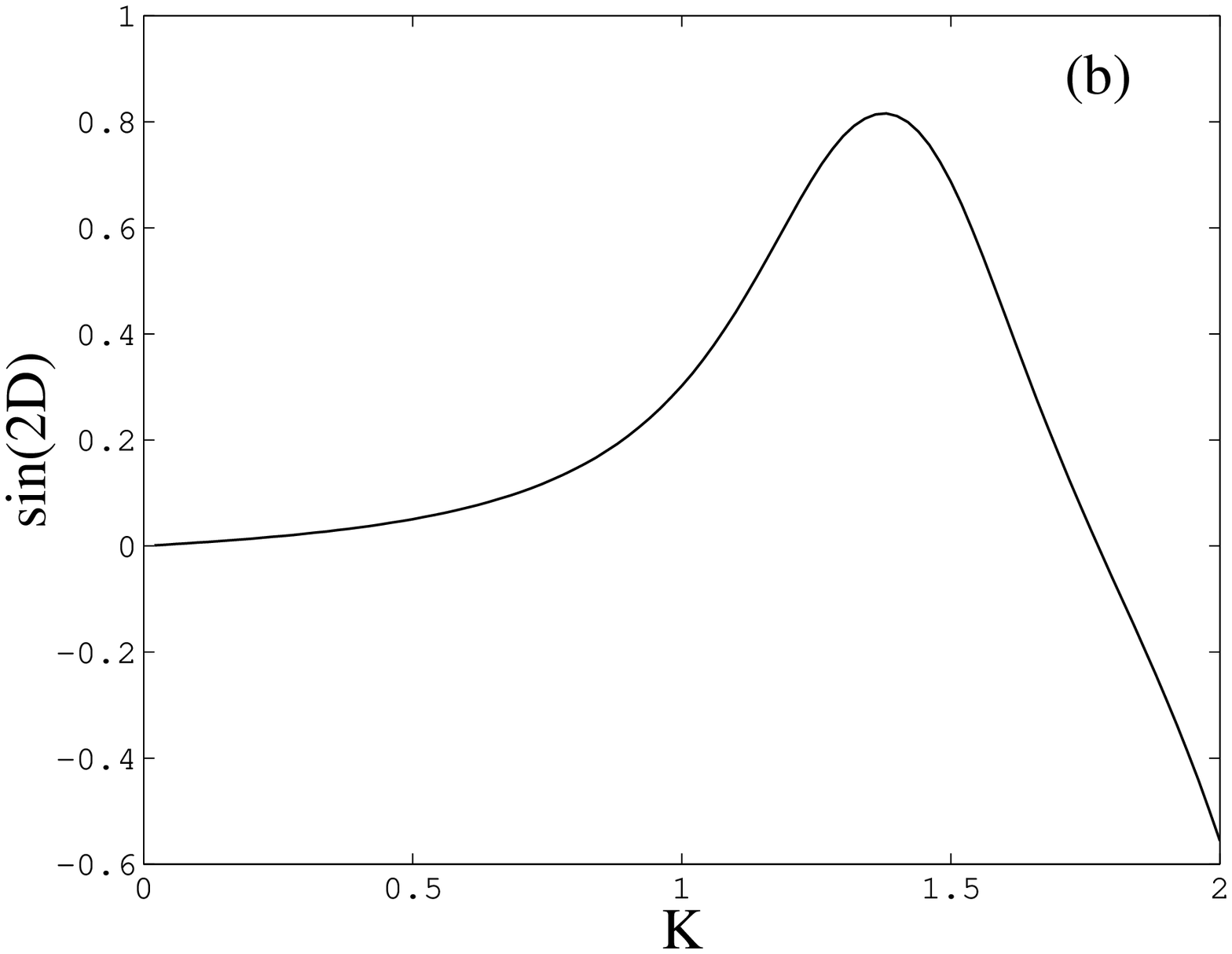}
\includegraphics[width=0.7\linewidth]{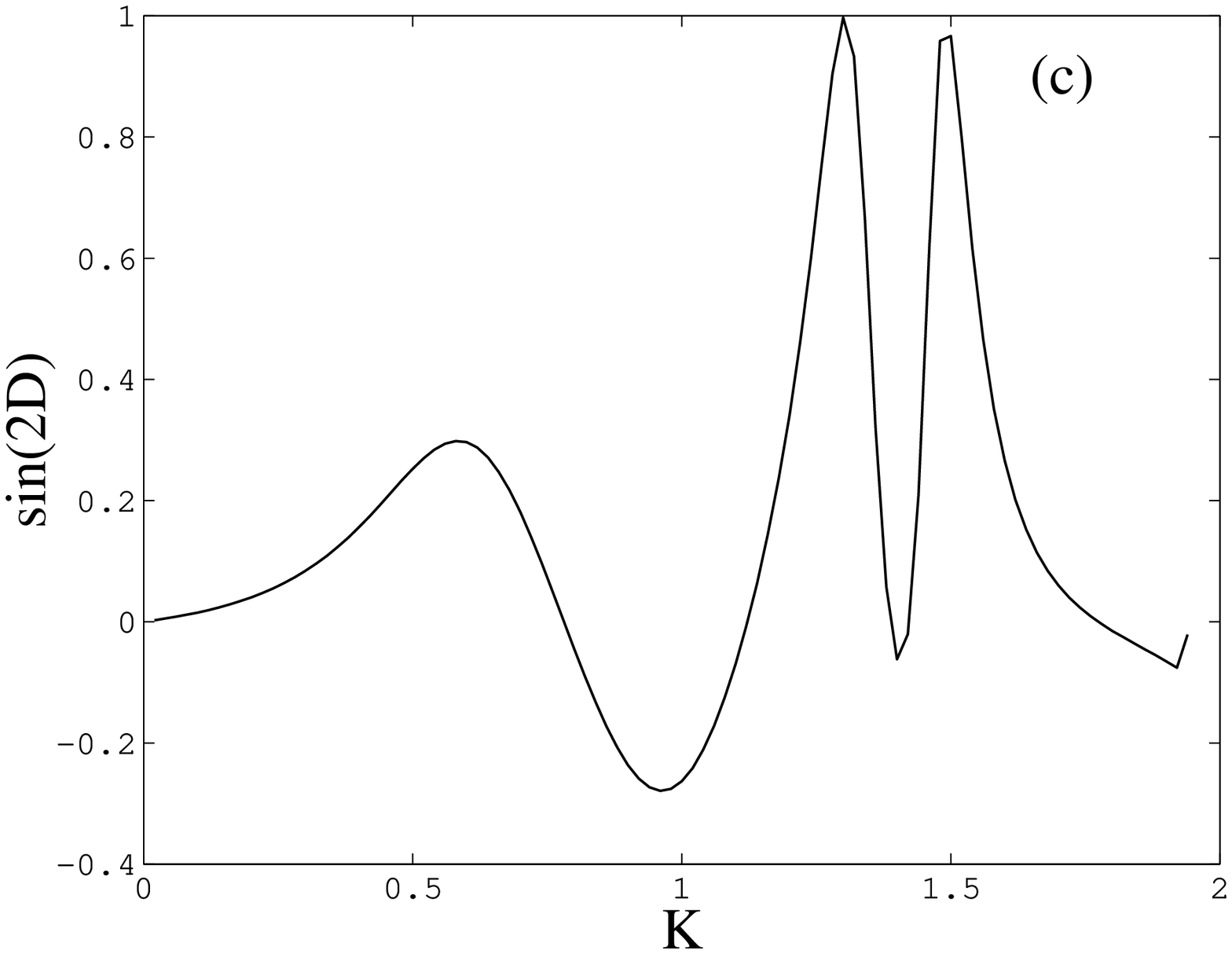}
\caption{The difference of optimal phases $D$ (see Eq.~(\ref{phase-dif-def}))
as a function
of the Chirikov parameter $K$ at a fixed number of kicks: (a) $n=2$, (b) $n=3$,
and (c) $n=4$. Everywhere $x_0=\pi$ and $\sigma=0.006$.}
\label{fig7ps}
\end{figure}

Basing our reasoning on a similar semiclassical picture, we examined the
stability of the time dependence of the optimal phase $\theta^*(t)$, which was
calculated quantum mechanically, against a small perturbation of the
initial position of the wave packet.
More precisely, we found the time dependence of the optimal phase
$\theta_1^*$ with the initial condition $x_0=\pi$ and, similarly,
$\theta_2^*(t)$ with the initial condition $x_0=\pi-0.05$. We denote the
difference of these phases by
\begin{equation}
\label{phase-dif-def}
D(t)=\theta_1^*(t)-\theta_2^*(t).
\end{equation}
Since $\theta^*(t)$ is periodic with a period $\pi$ (see Eq.~(\ref{17}),
it is natural to take $\sin(2D)$ as the quantity of interest, since m this
way we avoid breaks in the diagrams related to the periodicity of
$\theta^*(t)$. The dependence of $\sin(2D)$ on the Chirikov parameter $K$ for
different fixed numbers of kicks is depicted in
Figs.~\ref{fig7ps}a-\ref{fig7ps}c.
After two kicks (Fig.~\ref{fig7ps}a) the maximum value of $\mid\sin(2D)\mid$
does not exceed 0.035 at $K=2$. After three kicks ((Fig.~\ref{fig7ps}a)
the value of $D$ becomes significant at $K\gtrsim 1.2$. Finally,
after four kicks (Fig.~\ref{fig7ps}c) the process of measuring squeezing
becomes essentially unstable at $K\gtrsim 1$. Indeed, in these
conditions with a small perturbation of the initial position of the
wave packet, the difference of the optimum phases
reaches a value of order $\pi$. Such regime of squeezed states generation
was called \textit{unstable squeezing} in \cite{alekseev95}. As
Fig.~\ref{fig7ps} implies,
unstable squeezing is observed when chaos is strong.
On the other hand, for short time intervals and small values of $K$, the
squeezing can be already strong enough but yet rather stable.
\par
\begin{figure}
\includegraphics[width=1\linewidth]{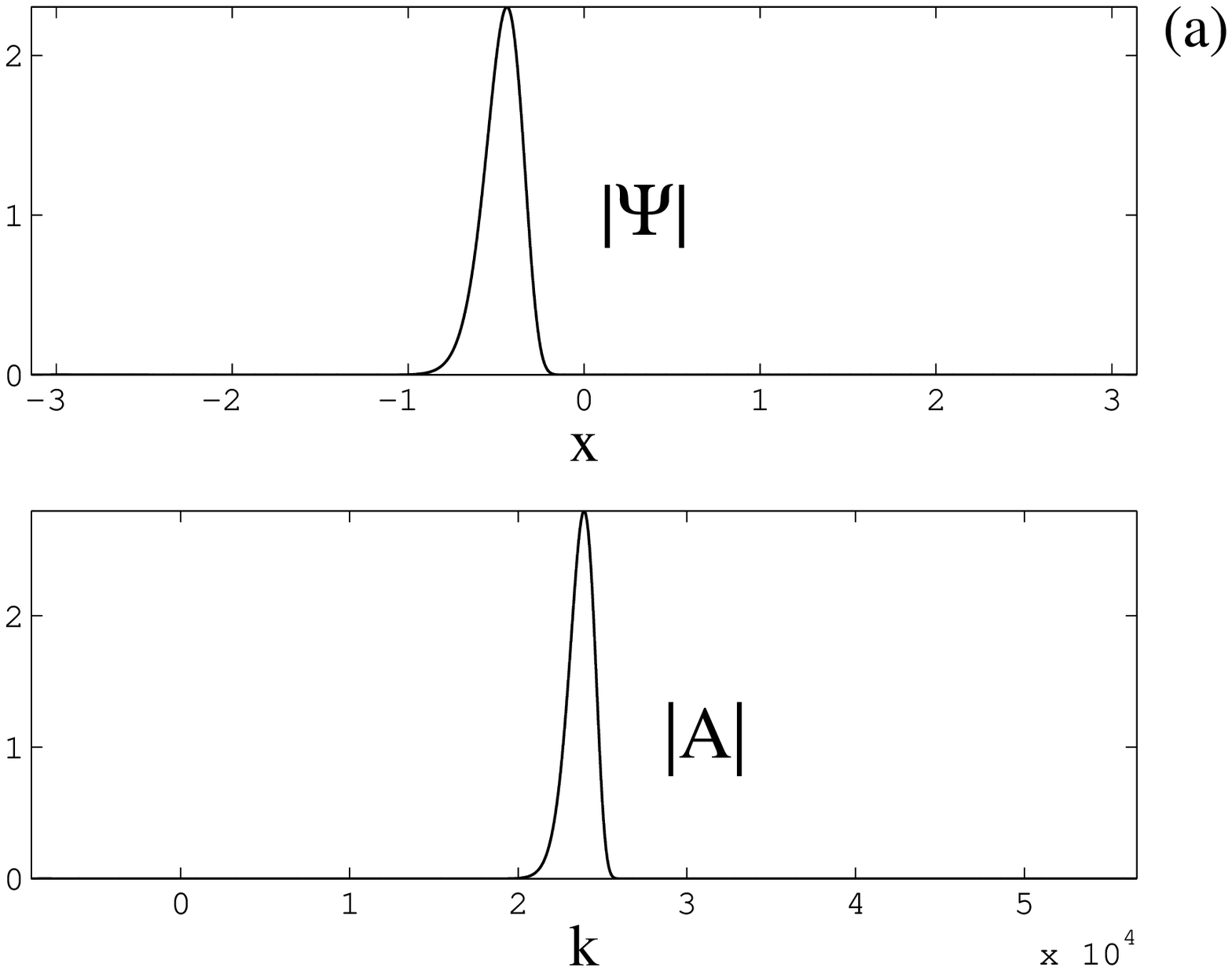}
\includegraphics[width=1\linewidth]{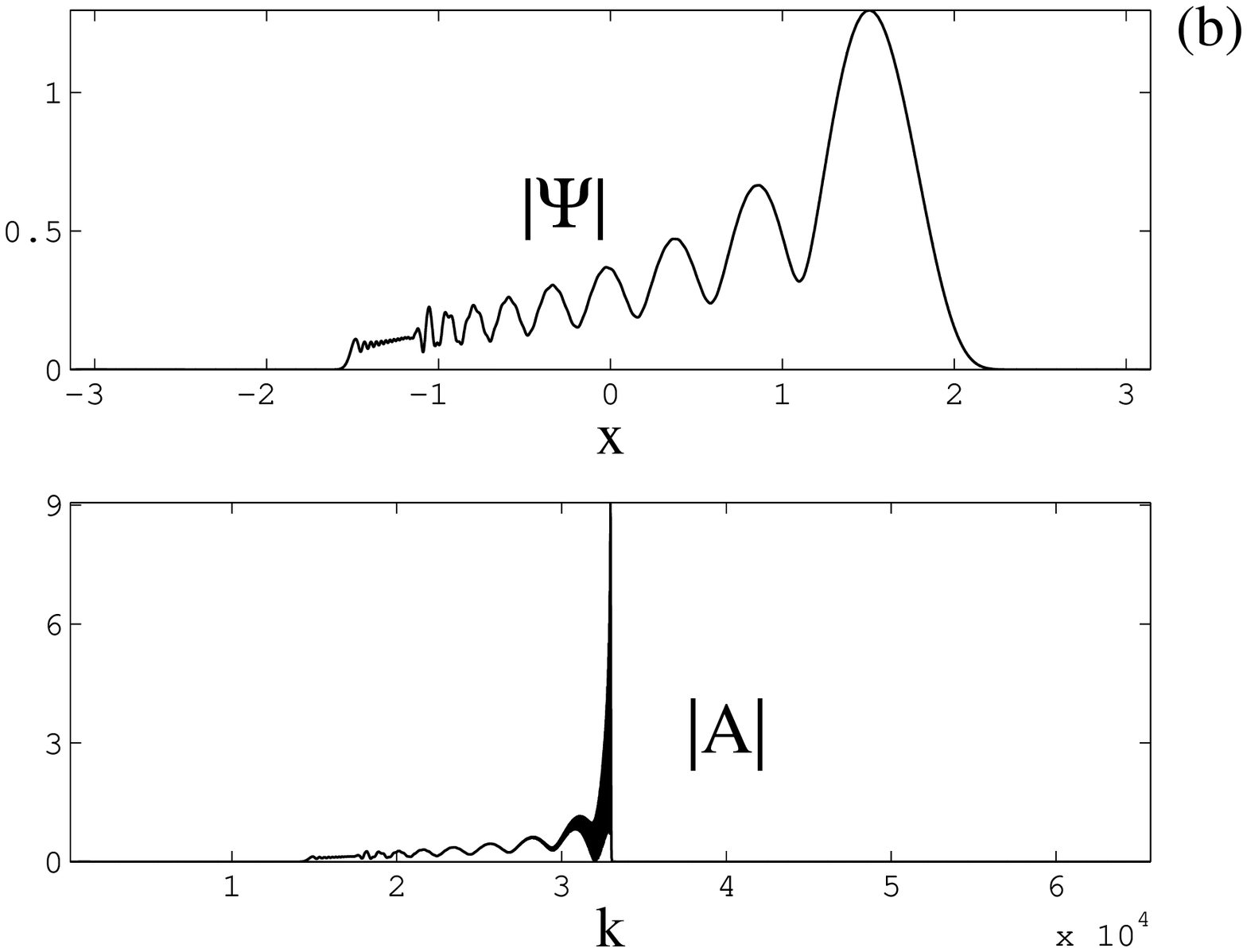}
\caption{Dependence of absolute value of the wave function
$\mid\Psi\mid$ on $x$ and dependence of the absolute value of the Fourier
transform $\mid A_k\mid$ on $k$ (see Eq.~(\ref{4})) at (a) $n=6$, (b) $n=18$.
Everywhere $K=1.2$, $x_0=\pi$ and $\sigma=0.006$.}
\label{fig8ps}
\end{figure}

To conclude this section we will briefly touch on the problem of the dynamics
of disintegration of coherent states in conditions of
chaos, a problem that is of interest by itself. Figures \ref{fig8ps}a and
\ref{fig8ps}b depict the dependence of $\mid\Psi\mid$ on $x$ and of
$\mid A\mid$ on $k$ (see
Eq. (\ref{4})).
Actually, Fig.~\ref{fig8ps} gives the shape of the wave function
in the coordinate and momentum representations for an
initially narrow wave packet with
$[\langle\delta x^2\rangle]^{1/2}(t=0)\equiv\sigma=0.006$ and
$[\langle\delta p^2\rangle]^{1/2}(t=0)=(1/12)\times 10^{-3}$.
The relatively small value $K=1.2$
makes it possible to examine the fairly long evolution of the wave packet
up to the point of its total
disintegration\footnote{In conducting numerical experiments in the dynamics
of the disintegration of wave packets we did not use the procedure
(described in Sec.~\ref{sec:Numerical_method}) of terminating the counting
process when the wave function becomes delocalized.
}.
After six kicks (Fig.~\ref{fig8ps}a) the wave packet spreads out significantly,
but on the whole retains its bell shaped structure.  What follows is a
disintegration of the packet into many small subpackets, with the
characteristic shape of the wave function depicted in Fig.~\ref{fig8ps}b
(after 18 kicks). Finally, very soon the wave function becomes so dissected
that even $2^{17}$ Fourier harmonics are
insufficient to describe the evolution correctly (for the data of
Fig.~\ref{fig8ps} this happens approximately at the 20th kick).
\par
Qualitatively, the same pattern of the evolution of the wave packet was
observed at higher values of $K$: first the
broadening of the wave packet, and then its rapid disintegration into many
very small subpackets. The
differences in packet disintegration for large values of $K$ in comparison
with the case $K\simeq 1$ (as in Fig.~\ref{fig8ps}) boil down to two
facts:(i) the ``swelling of the packet'' and the disintegration occur very
rapidly (it takes only several kicks to complete the process), and (ii)
the emerging subpackets are extremely small. Hence the process of
disintegration of wave packets in strong
chaos resembles an explosion. On the whole, the pattern being described
agrees well with the pattern obtained from the
analysis of the behavior of the Wigner function \cite{casati95}, although we
observed some anomalies. In particular, for fairly
narrow wave packets ($\sigma=4\times 10^{-3}$) we observed the disintegration
of the initial packet into two fairly large subpackets.
Ripples then appeared on the subpackets, and the two disintegrated into many
small packets.
A more detailed description of the disintegration of coherent states at chaos
will be a subject of our separate publication.

\section{Discussion and conclusion}
\label{sec:Disc}

Thus, in this work we numerically  study the dynamics
of generation of squeezed states in the
evolution of a Gaussian wave packet in the quasiclassical limit for the
model of a kicked quantum rotator. We show
that within the time interval where the packet is well-localized the
squeezing becomes stronger in the transition to
chaos. For strong chaos and in long time intervals, the squeezing process
becomes unstable. These results, obtained
through direct numerical simulation, are in good agreement with the results
derived using perturbation theory
for other models \cite{alekseev-preprint,alekseev95,alekseev97-98}.
\par
In the final stages of preparing the manuscript for press we became acquainted
with two recent papers \cite{xie} also devoted to
the problem of generating nonclassical states (squeezing and antibunching)
in the systems with quantum chaos. The authors of Ref. \cite{xie}
presented the results of numerical experiments on the dynamics of quadrature
squeezing in simple quantum
models that allow a transition to chaos in the classical limit:
the Lipkin-Meshkov-Glick model \cite{lipkin} and the Belobrov-
Zaslavsky-Tartakovskil model \cite{belobrov}. In contrast to our approach,
the authors of paper \cite{xie} were interested in the
long-time limit, when the wave packets are delocalized and this sense
the quantum-classical correspondence is
completely violated. They found that quadrature squeezing disappears at the
transition to quantum chaos, although to
some degree squeezing is always present in conditions of regular motion.
The nonzero squeezing in the conditions of quantum chaos and in
short-time limit has been also mentioned in \cite{xie},
but the enhance of squeezing was not observed, probably because in their
numerical simulations the quasiclassical parameter was not large enough:
only several hundred of quantum levels participated in the dynamics of the
system.
Thus, the results of \cite{xie} do not contradict ours but supplement them in
another limiting case, the limit of long times of motion.
\par
In conclusion we would like to make a remark concerning the possibility
of observation of squeezing
in quantum chaos on a time scale corresponding to a well-defined
quantum-classical correspondence. At present
essentially all experiments on light squeezing are done in the stationary
regime. Squeezing at the transition to quantum
chaos increases only over finite time intervals and in this sense is a
transient dynamical phenomenon. We hope that the development of effective
experimental methods for observing squeezed states of light in transient
dynamical regimes will also make it possible to observe the enhanced
squeezing at the transition to quantum chaos.
On the other hand, as noted in the Introduction, it is much simpler to realized
the quantum kicked rotator model in atomic optics \cite{atom-opt-exp}.
Moreover, it is much simpler to observe transient dynamical regimes in
experiments with cold atoms. Hence we believe that atomic optical systems
have great potential for observations of squeezed states at transition to
quantum chaos.
\par
We are grateful to Andrey Kolovsky and Jan Pe\v{r}ina for discussions and
to Boris Chinkov for support and drawing our attention to his
work \cite{chirikov-conf}.

\end{document}